\documentclass[
prd
,showpacs,amssymb,superscriptaddress,aps
]{revtex4}
\usepackage{graphicx}
\usepackage{color}
\input{epsf}

\usepackage{amsmath,amssymb}
\usepackage{bm}
\usepackage{times}
\usepackage{ulem}

\usepackage{hyperref}
\hypersetup{
    colorlinks=true,
    linkcolor=blue,
    filecolor=magenta,      
    urlcolor=cyan,
    citecolor=blue
}
\newcommand{\dalm}{\kern1pt\vbox{\hrule height 0.9pt\hbox{\vrule width 0.9pt
\hskip 2.5pt\vbox{\vskip 5.5pt}\hskip 3pt\vrule width 0.3pt}\hrule height 0.3pt}
\kern1pt}

\newcommand{\gsim}{\, \raisebox{-0.8ex}{$\stackrel{\textstyle >}{\sim}$ }}
\newcommand{\lsim}{\, \, \raisebox{-0.8ex}{$\stackrel{\textstyle <}{\sim}$ }}

\begin{document}



\title{Universal relation for supernova gravitational waves}

\author{Hajime Sotani}
\email{sotani@yukawa.kyoto-u.ac.jp}
\affiliation{Astrophysical Big Bang Laboratory, RIKEN, Saitama 351-0198, Japan}
\affiliation{Interdisciplinary Theoretical \& Mathematical Science Program (iTHEMS), RIKEN, Saitama 351-0198, Japan}

\author{Tomoya Takiwaki}
\affiliation{Division of Science, National Astronomical Observatory of Japan, 2-21-1 Osawa, Mitaka, Tokyo 181-8588, Japan}
\affiliation{Center for Computational Astrophysics, National Astronomical Observatory of Japan, 2-21-1 Osawa, Mitaka, Tokyo 181-8588, Japan}

\author{Hajime Togashi}
\affiliation{Department of Physics, Tohoku University, Sendai 980-8578, Japan}

\date{\today}

\begin{abstract}
Using the numerical simulation data for two-dimensional core-collapse supernova, we examine the protoneutron star (PNS) asteroseismology with the relativistic Cowling approximation. As shown in the previous study, the gravitational wave signals appearing in the numerical simulation can be well identified with the gravity (fundamental) oscillation in the early (later) phase before (after) the avoided crossing between the gravity and fundamental oscillations. On the other hand, the time evolution of supernova gravitational waves strongly depends on the PNS models, such as the progenitor mass and the equation of state for dense matter. Nevertheless, we find that the fundamental and gravity mode frequencies according to the gravitational wave signals appearing in the numerical simulations can be expressed as a function of the protoneutron star average density independently of the PNS  models. Using the average density, we derive the empirical formula for supernova gravitational wave frequency. In addition, we confirm that the dependence of the PNS surface density on the PNS average density is almost independent of the PNS models and also discuss how the different treatment of the non-uniform matter in the equation of state affects the observables.  
\end{abstract}

%
\maketitle


\section{Introduction}
\label{sec:I}

Direct detection of gravitational waves starts the era of gravitational wave astronomy and adds as a new messenger to electromagnetic waves and neutrinos with the observation of GW170817, which is the gravitational wave event from the binary neutron star merger \cite{GW6,EM}.
Now, we become able to use not only electromagnetic waves and neutrinos but also gravitational waves as a tool for probing astronomical phenomena. The observation of electromagnetic waves is complementary important in gravitational wave astronomy, especially for determining the source position, because the angular resolution with gravitational wave detectors is poor, spatially if only a limited set of detectors are used.
For the case of the GW170817, searches within the sky location area provided by the gravitational wave detectors actually were crucial to finding the faint kilonova by electromagnetic observations, while it was possible to improve the sky location of the gravitational wave event, assuming that it was at the location of the kilonova.
The third observing run (O3) by LIGO and Virgo has already finished, where KAGRA finally joined this collaboration network at the end of O3, and these gravitational wave detectors are now preparing for O4, which may start after August 2022. During the O3, the gravitational waves from two merger events composed of a neutron star with a black hole, which are the last piece in the compact binary merger event, have also been detected \cite{BHNS}. Owing to the development of detectors' sensitivity, the observable distance is growing and the number of events is expected to significantly increase.

The next frontier in gravitational wave astronomy after the compact binary mergers must be the (galactic) core-collapse supernova, which accompanies the death of massive star. Since, compared to the compact binary mergers, the supernova is almost spherically symmetric, the radiation energy of gravitational waves may not be so strong, which leads to a difficulty of detection from the supernova explosion quite far from us. Considering the event rate of supernova, e.g, $\sim1$ per century in our galaxy, the detection of the gravitational wave from such a system seems to be quite rare. Nevertheless, once one would detect it, one could simultaneously detect not only the gravitational waves but also the electromagnetic waves and neutrinos. This must be a good opportunity for understanding the supernova explosion mechanism, and we have to wait for such an opportunity with thoroughgoing conditions.

The gravitational waves from supernova explosions have been mainly studies via numerical simulations (e.g., \cite{Murphy09,MJM2013,Ott13,CDAF2013,Yakunin15,KKT2016,Andresen16,Richers2017,Takiwaki2017,OC2018,RMBVN19,VBR2019,PM2020,Andersen21,Pajkos21,Shibagaki21}). These simulations tell us the existence of the gravitational wave signals, whose frequency increases from a few hundred hertz up to a few kilohertz in $\sim 1$ second after corebounce. This ramp up signal is originally considered as a result of the Brunt-V\"{a}is\"{a}l\"{a} frequency at the surface of the protoneutron star (PNS), the so-called surface gravity ($g$-) mode \cite{MJM2013,CDAF2013}. Meanwhile, since the Brunt-V\"{a}is\"{a}l\"{a} frequency determined with local stellar properties strongly depends on the radial position inside the star, the gravitational wave signals appearing in the numerical simulations may come from specific global oscillations of the dense region. In fact, the signals can be identified with the frequency of the PNS fundamental ($f$-) mode \cite{MRBV2018,SKTK2019,ST2020b}, where the signals in very early phase just after corebounce is identified with the $g_1$-mode oscillations \cite{ST2020b}, or with the $g$-mode like oscillations of the region inside the shock radius \cite{TCPF2018,TCPOF2019a,TCPOF2019b}. Anyway, the identification of the gravitational wave signals with a specific oscillation mode must be important for understanding the physics behind the gravitational wave radiations.

In order to study the gravitational waves from the core-collapse supernovae, the perturbative analysis is considered as another approach. In general, the oscillation frequencies strongly depend on the interior properties of the source object, where the excited frequencies respectively correspond to the input physics \cite{KS1999}. So, as an inverse problem, one could extract the interior properties, when one would detect the frequency. This technique is known as asteroseismology, which is similar to seismology on the Earth and helioseismology on the Sun. For compact objects, asteroseismology on cold neutron stars is extensively studied up to now. For instance, by identifying the frequencies of quasi-periodic oscillations observed in the afterglow following the giant flares with the crustal torsional oscillations, the crust properties in neutron stars are constrained \cite{SA09,GNHL2011,SNIO2012,Deibel14,Gabler2018}. In a similar way, once one would detect the oscillation frequencies in gravitational waves from neutron stars, one might constrain the properties of the neutron stars, such as the mass, radius, equation of state (EOS), and so on (e.g., \cite{AK1996,AK1998,STM2001,SH2003,SYMT2011,PA2012,DGKK2013,Sotani2020}).

On the other hand, the studies about asteroseismology on PNSs are gradually increasing (e.g., \cite{MRBV2018,SKTK2019,ST2020b,TCPF2018,TCPOF2019a,TCPOF2019b,FMP2003,Burgio2011,FKAO2015,ST2016,Camelio17,SKTK2017,SS2019,WS2019,WS2020,Mezzacappa20,ST2020a,ST2020c}). This may be partially thanks to progress of the numerical simulation of core-collapse supernovae, which can provide the background PNS models for making a linear analysis. In these studies, using the numerical data obtained by the simulation of core-collapse supernovae, first the spherically symmetric background models are prepared by averaging in the angular direction, if the multi-dimensional simulation has been done. With using such a background model, mainly two different approaches have been done in the PNS asteroseismology. The first approach is that the PNS surface is defined by a specific surface density, e.g., $\rho_s=10^{11}$\,g/cm$^3$, and one considers the characteristic oscillations inside the PNS model, where the boundary condition imposed at the PNS surface is assumed to be
 the vacuum boundary condition, i.e., the Lagrangian perturbation of pressure should be zero \cite{MRBV2018,SKTK2019,ST2020b,ST2016,FMP2003,Burgio2011,FKAO2015,Camelio17,SKTK2017,SS2019},
 even though the matter still exists outside the PNS surface selected here. With this approach, one can classify the oscillation modes with the number of radial nodes in the eigenfunctions according to the standard asteroseismology, while there is uncertainty in the selection of the surface density. But, it has been shown that at least the $f$- and $g_1$-mode frequencies in the later phase (e.g., since $\sim$ 300\,msec after core bounce) are almost independent of the selection of surface density \cite{MRBV2018,ST2020b}. In the second approach, one adopts the background data up to the shock radius and imposes the boundary condition that the position of the shock radius is fixed during the oscillations \cite{TCPF2018,TCPOF2019a,TCPOF2019b}. With the second approach, by definition  one can uniquely fix the position of the outer boundary, but one has to reclassify the oscillation modes because the eigenvalue problem to solve is mathematically different from the standard asteroseismology due to the different boundary condition at the outer boundary. In this study, we simply adopt the first approach, because we mainly focus on the $f$- and $g_1$-mode frequencies, which correspond to the gravitational wave signals appearing in the numerical simulation~\cite{ST2020b}.

Nonetheless, since the $f$- and $g_1$-mode frequencies generally depend on the PNS models (e.g., \cite{Andersen21}), it may be difficult to extract a kind of physical properties of PNSs via direct observation of the supernova gravitational waves. To solve this difficulty, the universal relation between the $f$- and $g_1$-mode frequencies and PNS properties, which is independent of the PNS models, must be quite important, if it exists. In fact, with the same motivation the universal relation of the oscillation frequency as a function of the PNS surface gravity is proposed in Refs.~\cite{TCPOF2019b,Bizouard2021}. 
On the other hand, we alternatively proposed another fitting formula for the supernova gravitational waves as a function of the PNS average density in Ref.~\cite{ST2020b}, which fits well the gravitational wave frequencies just after the corebounce. However, we did not discuss how universal our fitting formula can predict the supernova gravitational wave frequency. Thus, in this study we mainly focus on the discussion for the universality of our fitting formula. In addition, we will confirm another universal relation between the PNS surface gravity and PNS average density, which is originally found in Ref.~\cite{ST2020a}. Furthermore, we also discuss how the observables, such as the PNS mass, radius, and gravitational waves, depend on the different treatment of the non-uniform matter in PNS.

This paper is organized as follows. In Sec. \ref{sec:PNSmodel}, we describe the PNS models considered in this study. In Sec. \ref{sec:GW}, we show the eigenfrequencies of gravitational waves from the PNS with the relativistic Cowling approximation and show the universal relation of gravitational wave frequencies appearing in the numerical simulations as a function of the PNS average density. Then, we make a conclusion in Sec. \ref{sec:Conclusion}.

\section{PNS Models}
\label{sec:PNSmodel}

In order to identify the gravitational wave signals appearing in the core-collapse supernova simulations with a  specific PNS oscillation frequency, we examine asteroseismology on the PNSs. For this purpose, first we have to prepare the PNS model as a background model, on which the linear analysis is done. As in the previous our studies \cite{ST2020a,ST2020b,ST2020c}, such background PNS models are provided by the two-dimensional (2D) neutrino radiation hydrodynamic simulations with {\small 3DnSNe} code, which has been already applied for several studies \cite{takiwaki2016,oconnor2018,kotake2018,matsumoto2020}. In this numerical code, the neutrino transport is solved by adopting isotropic diffusion source approximation \cite{liebendoerfer2009,takiwaki2014}, where we adopt the same procedure for the set of neutrino reactions as in Ref.~\cite{oconnor2018}. The numerical simulations have been done with the resolution in the spherical polar grid of $512\times 128$, where the radial grid covers up to $5000$\,km from the center, adopting the $20M_\odot$ progenitor model given in Ref.~\cite{WH2007}.

With respect to the EOS for high density region, we adopt four different EOSs in this study, i.e., DD2, SFHo, TGLD, and TGTF. For uniform matter, DD2 is constructed based on the relativistic mean field approach with the density-dependent parametrization \cite{DD2}, while SFHo is constructed with another parametrization that is fitted to the neutron star mass and radius constrained from the x-ray observations  \cite{SFHo}. Non-uniform matter in both EOSs are described by the nuclear statistical equilibrium (NSE) model, where nuclei are regarded as classical particles. TGTF \cite{Togashi17} and TGLD \cite{Furusawa17} basically have the same EOS properties for uniform matter, which are constructed via variational method with AV18 and UIX potential. The difference is how to construct a non-uniform matter density region, i.e., TGTF is adopted single nucleus approximation (SNA) with Thomas-Fermi model, while TGLD is constructed by the NSE treatment with liquid drop model. The EOS parameters adopted in this study are listed in Table \ref{tab:EOS}, where $K_0$ and $L$ are respectively incompressibility of the symmetric nuclear matter and the density gradient of the symmetry energy at the saturation point, while $\eta$ given by the combination of $K_0$ and $L$ as $\eta\equiv (K_0 L^2)^{1/3}$ is a suitable parameter for expressing the low-mass (cold) neutron stars \cite{SIOO14}. Considering the constraint from GW170817 on the radius of a $1.4M_\odot$ neutron star \cite{Annala18}, i.e., $R\le 13.6$\,km, $\eta$ may be also constrained as $\eta\lsim 125$ MeV. In the same table, the maximum mass of a corresponding cold neutron star is also listed.

In addition to the above PNS models, for reference we also consider the results in Ref.~\cite{ST2020b}, which has been studied with completely different PNS model. That is, the $2.9M_\odot$ helium star (He2.9) given in Ref.~\cite{He29} is adopted as a progenitor model and LS220 \cite{LS220} is adopted for the EOS, with which the 2D numerical simulation has been done. Anyway, in order to prepare the spherically symmetric background model, the numerical data obtained by 2D simulations is averaged in the angular ($\theta$) direction. In this study, we especially set the PNS surface density to be $\rho_s=10^{11}$\,g/cm$^3$.

\begin{table}
\caption{EOS parameters, $K_0$, $L$, and $\eta$, adopted in this study. In the rightmost column, the maximum mass, $M_{\rm max}/M_\odot$, for the cold neutron star constructed with each EOS is listed. For reference, the properties for LS220 are also shown. } 
\label{tab:EOS}
\begin {center}
\begin{tabular}{ccccc}
\hline\hline
EOS & $K_0$ (MeV) & $L$ (MeV) & $\eta$ (MeV) & $M_{\rm max}/M_\odot$   \\
\hline
DD2 & 243 & 55.0  & 90.2  & 2.42  \\
SFHo & 245 & 47.1 & 81.6 & 2.06  \\
TGLD/TGTF & 245  & 38.7  & 71.6 & 2.21   \\ 
LS220 & 220 & 73.8 & 106.2 & 2.05 \\
\hline \hline
\end{tabular}
\end {center}
\end{table}

\begin{figure}[tbp]
\begin{center}
\includegraphics[scale=0.5]{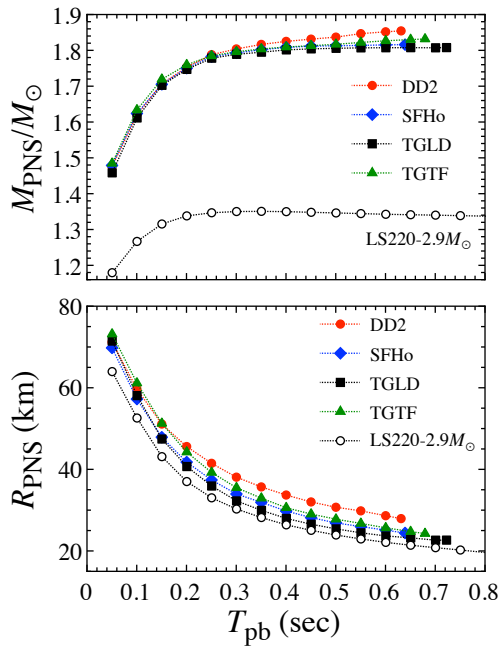} 
\end{center}
\caption{
Time evolution of the PNS mass (top) and radius (bottom) for various models. The surface density is chosen to be  $10^{11}$\,g/cm$^3$.
}
\label{fig:MRt}
\end{figure}

\begin{figure}[tbp]
\begin{center}
\includegraphics[scale=0.5]{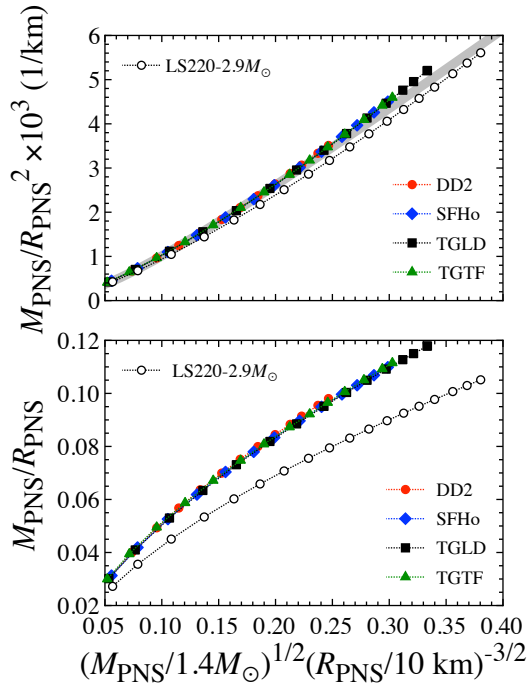} 
\end{center}
\caption{
The PNS surface gravity (top) and compactness (bottom) are shown as a function of the square root of the normalized PNS average density for various models. The thick-solid line in the top panel is the fitting formula given by Eq.~(\ref{eq:surfgrav}).
}
\label{fig:surfgrav}
\end{figure}

In Fig.~\ref{fig:MRt} we show the PNS mass (top) and radius (bottom) evolution for various PNS models. One can clearly observe that the difference of the progenitor mass significantly changes the PNS mass evolution, but it affects the PNS radius evolution not so much. In addition, we find a deviation between the results with TGTF and TGLD, i.e., the PNS mass and radius evolution depends on the treatment of non-uniform matter (SNA or NSE). This dependence will discuss later together with that of the gravitational wave frequencies.

Additionally, in Fig.~\ref{fig:surfgrav}, we show the surface gravity, $M_{\rm PNS}/R_{\rm PNS}^2$, in the top panel and PNS compactness, $M_{\rm PNS}/R_{\rm PNS}$, in the bottom panel as a function of the square root of the normalized PNS average density, $(M_{\rm PNS}/1.4M_\odot)^{1/2}(R_{\rm PNS}/10\ {\rm km})^{-3/2}$, for various PNS models. In the both panels, the PNSs evolve from the bottom-left to the top-right. From this figure, in particular, we can confirm that the surface gravity is quite weak dependence on the square root of the normalized PNS average density, which is originally pointed out in Ref.~\cite{ST2020a}. With the results, we derive the fitting formula as
\begin{equation} 
  M_{\rm PNS}/R_{\rm PNS}^2 \ ({\rm km^{-1}}) = \left[-1.985 -0.465\ln(x) + 19.247 x\right]\times 10^{-3}, 
  \label{eq:surfgrav}
\end{equation}
where $x$ denotes the square root of the normalized PNS average density, i.e., 
\begin{equation}
  x \equiv \left(\frac{M_{\rm PNS}}{1.4M_\odot}\right)^{1/2} \left(\frac{R_{\rm PNS}}{10\ {\rm km}}\right)^{-3/2},
\end{equation}
which is shown in the top panel of Fig.~\ref{fig:surfgrav} with the thick-solid line. To be honest, currently we can not understand why this relationship appears in the PNS models, because this type of the relation can not be seen in cold neutron stars, but this feature may be a key for understanding the physics of PNSs.

\section{gravitational wave asteroseismology}
\label{sec:GW}

On the PNS models discussed in the previous section, we make a linear analysis, where we simply adopt the relativistic Cowling approximation in this study, i.e., the metric perturbations are neglected during the fluid oscillations. The perturbation equations are derived by linearizing the energy conservation law, which finally become the ordinary differential equations for the Lagrangian displacement of fluid element. By imposing appropriate boundary conditions, the problem to solve becomes an eigenvalue problem with respect to the eigenvalue of $\omega$, which directly corresponds to the eigenfrequency, $f$, via $f=\omega / (2\pi)$. The imposed boundary conditions are the regularity condition at the center and the condition that the Lagrangian perturbation of pressure should be zero at the PNS surface. The concrete perturbation equations and boundary conditions are the same as shown in Ref.~\cite{SKTK2019}. In this study, we focus on only the $\ell=2$ oscillation modes, because they are considered to become energetically dominant in the gravitational wave emission. 
We note that one can qualitatively discuss the behavior of gravitational wave frequency even with the relativistic Cowling approximation, but the frequencies with the approximation deviate $\sim 20 \%$ from those without the approximation \cite{ST2020c}.

\begin{figure}[tbp]
\begin{center}
\includegraphics[scale=0.5]{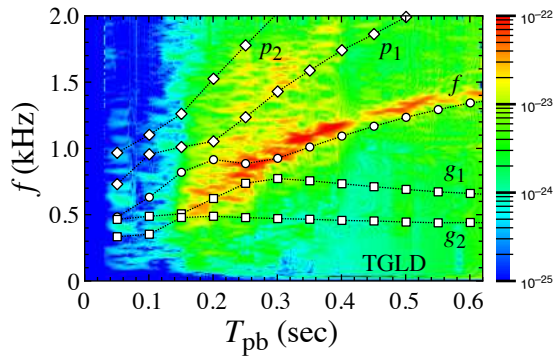} 
\end{center}
\caption{
Comparing the gravitational wave signals appearing in the numerical simulation (background contour) to the PNS frequencies (open-marks) determined by solving the eigenvalue problem for the PNS model with TGLD.
}
\label{fig:comp-TGLD}
\end{figure}

\begin{figure}[tbp]
\begin{center}
\includegraphics[scale=0.5]{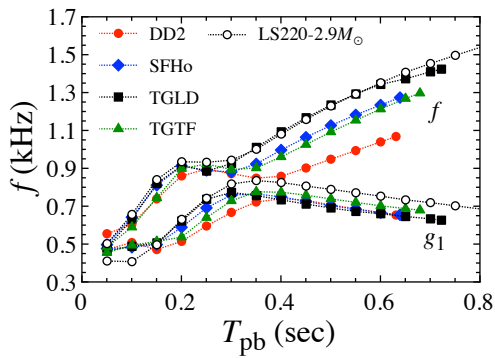} 
\end{center}
\caption{
The $f$- and $g_1$-mode frequencies for various PNS models are shown as a function of the postbounce time.
}
\label{fig:ft}
\end{figure}

In Fig.~\ref{fig:comp-TGLD} we show the PNS oscillation frequencies determined by solving the eigenvalue problem with open marks on the contour, which denotes the gravitational wave signals appearing in the numerical simulation, for the PNS model with TGLD (see in appendix \ref{sec:appendix_1} for the other PNS models), focusing on only the $f$-, $g_i$-, and $p_i$-mode frequencies with $i=1,2$. Here, the gravitational wave signals are calculated with the same procedure as in Ref.~\cite{Marek06}, using the numerical data obtained by simulations. From this figure, as in Ref.~\cite{ST2020b}, one can obviously see that the gravitational wave signals in numerical simulation is identified by the $g_1$-mode ($f$-mode) oscillation from the PNS before (after) the avoided crossing between the $f$- and $g_1$-mode. In Fig.~\ref{fig:ft}, we also plot the time evolution of the $f$- and $g_1$-mode frequencies for various PNS models. As in Ref.~\cite{Andersen21}, one can observe that the time evolution of the gravitational waves strongly depends on the PNSs models (see also Fig.~\ref{fig:comp}).

\begin{figure}[tbp]
\begin{center}
\includegraphics[scale=0.5]{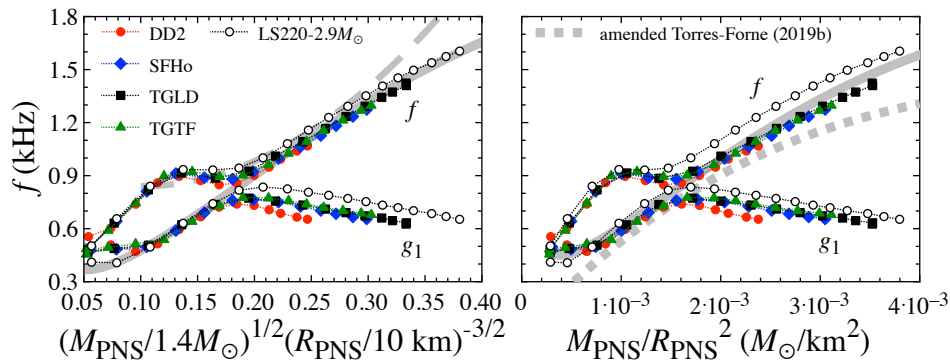} 
\end{center}
\caption{
In the left panel, the $f$- and $g_1$-mode frequencies for various PNS models are shown as a function of the square root of the normalized PNS average density. The thick-solid line is the fitting formula for the $g_1$-mode ($f$-mode) frequency before (after) the avoided crossing between the $f$- and $g_1$-modes, which is given by Eq.~(\ref{eq:universal}), while thick-dashed line denotes the empirical formula derived in Ref.~\cite{SS2019} (Eq.~(\ref{eq:ss2019})). On the other hand, in the right panel the same frequencies shown in the left panel are shown as a function of the surface gravity, where the thick-solid line denotes the fitting formula given by Eq.~(\ref{eq:fsurf}).
The thick-dotted line is the universal relation derived in Ref.~\cite{TCPOF2019b}, but it is amended \cite{private}.
}
\label{fig:fx-universal}
\end{figure}

On the other hand, in the left panel of Fig.~\ref{fig:fx-universal}, we show the $f$- and $g_1$-mode frequencies as a function of the square root of the normalized PNS average density, $x$. From this figure, the $f$- and $g_1$-mode frequencies according to the gravitational wave signals appearing in the numerical simulation are well fitted, such as 
\begin{equation}
 f \ {\rm (kHz)} = -1.410 - 0.443 \ln(x) + 9.337x -6.714x^2, \label{eq:universal}
\end{equation}
independently of the PNS models. The predicted values from Eq.~(\ref{eq:universal}) are also plotted in the left panel of Fig.~\ref{fig:fx-universal} with the thick-solid line. That is, once one would detect the supernova gravitational waves, which could be the same as gravitational wave signals appearing in the numerical simulations, one can extract the evolution of the PNS average density by using Eq.~(\ref{eq:universal}). In the same figure, we also show the empirical relation for the $f$-mode frequency derived in Ref.~\cite{SS2019}, which is 
\begin{equation}
  f_f {\rm (kHz)} = 0.9733 - 2.7171x + 13.7809x^2, \label{eq:ss2019}
\end{equation}
with the thick-dashed line. We remark that this relation is obtained for the case of the failed supernova with general relativistic simulation, i.e., the PNS considered in Ref.~\cite{SS2019} would eventually collapse to a black hole, 
focusing on the region of $(M_{\rm PNS}/1.4M_\odot)^{1/2}(R_{\rm PNS}/10\ {\rm km})^{-3/2}\gsim 0.1$. 
By comparing this empirical relation to the gravitational wave frequencies obtained in this study and the fitting formula given by Eq.~(\ref{eq:universal}), one can observe a significant deviation for the later phase. Unfortunately, we can not identify why this deviation comes from, but it may be because the dependence of the gravitational wave frequencies on the PNS average density in a black hole formation is simply different from that for the case of successful supernova, or it may comes from the treatment of the general relativistic effect in the simulation. On the other hand, it is also suggested that the gravitational wave frequency is expressed as a function of the PNS surface gravity in Refs.~\cite{TCPOF2019b,Bizouard2021}. In a similar way, we also show the gravitational wave frequencies for various PNS models as a function of the PNS surface gravity in the right panel of Fig.~\ref{fig:fx-universal}, where the thick-solid line denotes the fitting formula given by 
\begin{equation}
  f {\rm (kHz)} = -0.0752 -  0.2600\ln(\bar{x}_{3}) + 0.7446\bar{x}_{3} - 0.0600\bar{x}_{3}^2, \label{eq:fsurf}
\end{equation} 
where $\bar{x}_{3}$ denotes $\bar{x}/0.001$ and $\bar{x}$ is the PNS surface gravity defined by $\bar{x}\equiv M_{\rm PNS}/R_{\rm PNS}^2$ in the unit of $M_\odot$ km$^{-2}$.
For reference, we also show the universal relation derived in Ref.~\cite{TCPOF2019b} with the thick-dotted line, where the standard deviation of the data is 76\,Hz. 
We note that the universal relation in Ref.~\cite{TCPOF2019b} had a missing factor and the amended relation is plotted in this figure \cite{private}, instead of the original relation.
Since we have already shown that the relation between the PNS surface gravity and the average density weakly depends on the PNS model, as shown in the top panel of Fig.~\ref{fig:surfgrav}, we expected that the gravitational wave frequencies corresponding to the signals in numerical simulation could be also expressed as a function of the PNS surface gravity almost independently of the PNS models. However, the gravitational wave frequencies seem to depend a little on the PNS models, if they are considered as a function of the PNS surface gravity. That is, to characterize the gravitational wave frequency, the PNS average density may be better than the surface gravity, based on the eigenfrequencies computed using the linearised theory with the Cowling approximation. Anyway, as shown in Figs. \ref{fig:comp-TGLD} and \ref{fig:comp}, the resultant eigenfrequencies systematically deviate from the gravitational wave signals appearing in the numerical simulation, which is the same order of magnitude as the differences seen between modes when using surface gravity.

\begin{figure}[tbp]
\begin{center}
\includegraphics[width=.6\linewidth]{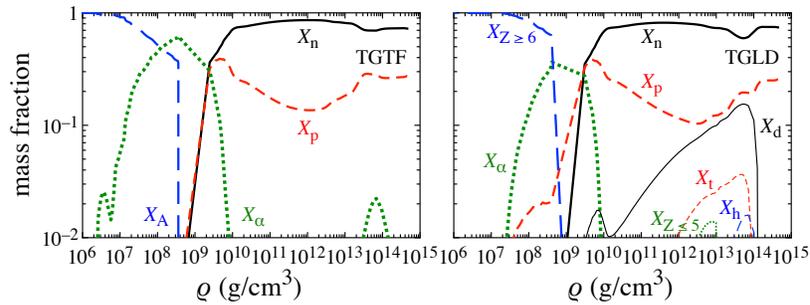} 
\end{center}
\caption{
Mass fraction for the models with TGTF (left panel) and TGLD (right panel) at $T_{\rm pb}=0.07$ sec. 
}
\label{fig:fraction}
\end{figure}

Furthermore, we discuss the impact on the observables due to the different treatment of non-uniform matter in EOS. That is, as mentioned before, TGTF and TGLD are constructed with the same nuclear properties for uniform matter, but the treatment in non-uniform matter  is different from each other.  This difference hardly affects the cold neutron star properties \cite{Togashi17,Furusawa20}, while we find that it strongly affects the evolution of the PNS properties and gravitational wave frequencies in this study.
The treatment of non-uniform matter affects the nuclear composition at the PNS surface, as in Fig.~\ref{fig:fraction}, where the left and right panels correspond to the models with TGTF and TGLD at $T_{\rm pb} = 0.07$\,sec. In the figure, the mass fractions of the neutron, proton, alpha particle, and representative single nucleus for the models with TGTF are shown as $X_n$, $X_p$, $X_\alpha$, and $X_{\rm A}$, while those of the neutron, proton, deuteron, triton, helion, alpha particle, other light nuclei with $Z\le 5$, and nuclei with $Z\ge 6$ for the models with TGLD are shown as $X_n$, $X_p$, $X_{d}$, $X_{t}$, $X_h$, $X_\alpha$, $X_{\rm Z \le 5}$, and $X_{\rm Z \ge 6}$, respectively. From this figure, one can observe that for the models with TGLD not only the neutron, proton, and alpha particle, but also various nuclei, deuteron, triton, and helion can appear even inside the PNS region, which corresponds to the density region for $\rho\ge 10^{11}$ g/cm$^3$.

\begin{figure}[tbp]
\begin{center}
\includegraphics[width=.6\linewidth]{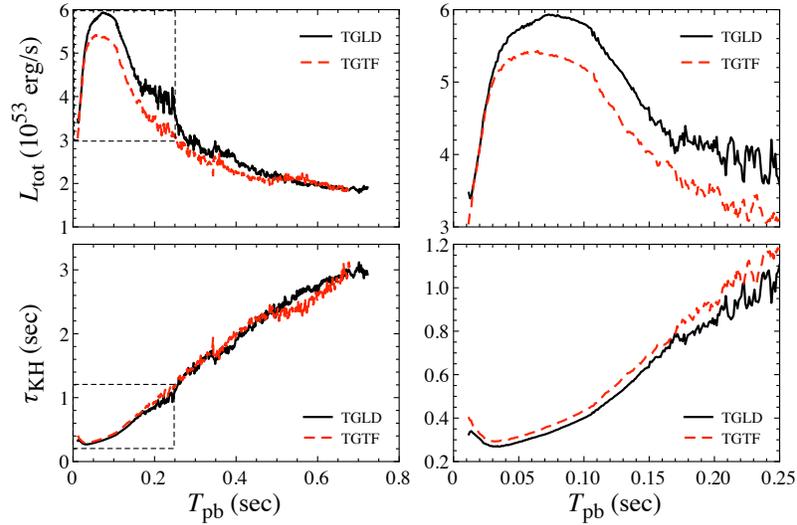}
\end{center}
\caption{
Total neutrino luminosity at the PNS surface, $L_{\rm tot}$, and the Kelvin-Helmholtz timescale, $\tau_{\rm KH}$, given by Eq.~(\ref{eq:tauKH}) are shown as a function of $T_{\rm pb}$ in the top and bottom panels, respectively. The right panels are just extended view of the region enclosed by dashed line in the left panels. 
}
\label{fig:pns-cooling}
\end{figure}

The difference between the results with TGTF and TGLD may be understood as follows.
The nuclear composition changes the strength of the neutrino cooling. At the PNS surface, the absorption and scattering by nucleon are the dominant opacity source. Since we especially treat light elements as alpha particles in our hydrodynamic simulations, the appearance of the deuteron and other light elements in TGLD reduces the fraction of nucleon and hence their opacity at the PNS  surface. The top panel of Fig.~\ref{fig:pns-cooling} shows the total neutrino luminosity at the PNS  surface given by $L_{\rm tot}=L_{\nu_e}+L_{\bar{\nu}_e}+4L_{\nu_X}$, where $\nu_e$, $\bar{\nu}_e$ and $\nu_X$ represent electron type, anti-electron type, and heavy lepton type neutrinos, respectively. 
The higher neutrino luminosity at TGLD is consistent with their fast contraction than TGTF (see Fig.~\ref{fig:MRt}).
We also evaluate the Kelvin-Helmholtz timescale, which is roughly the timescale of the contraction of the stars \cite{Kippenhahn12}, given by 
\begin{equation}
    \tau_{\rm KH} = |E_{\rm g}|/L_{\rm tot}. \label{eq:tauKH}
\end{equation}
In this equation, $E_{\rm g}$ is the the gravitational binding energy of the PNS defined by $E_{\rm g}=\frac{1}{2}\int  \rho \Phi {\rm d}V$, where $\Phi$ is the gravitational potential and the volume integration is performed in PNS. The bottom panel of Fig.~\ref{fig:pns-cooling} shows this  timescale. Reflecting the larger luminosity of TGLD, the timescale is smaller in TGLD in early phase. The range of the timescale, which is from 0.3\,sec to several seconds, is roughly consistent with the typical value shown in Ref.~\cite{Scheck06} (see also  Refs.~\cite{Nakazato19,Nakazato20,Nakazato21} for later epoch).
Note that if the emission and absorption process of deuteron and other light elements are considered as in Refs.~ \cite{Nagakura19,Fischer20}, which increases the opacity of PNS, the difference of the timescale between TGLD and TGTF could be smaller than that shown in this study.

\begin{figure}[tbp]
\begin{center}
\includegraphics[scale=0.5]{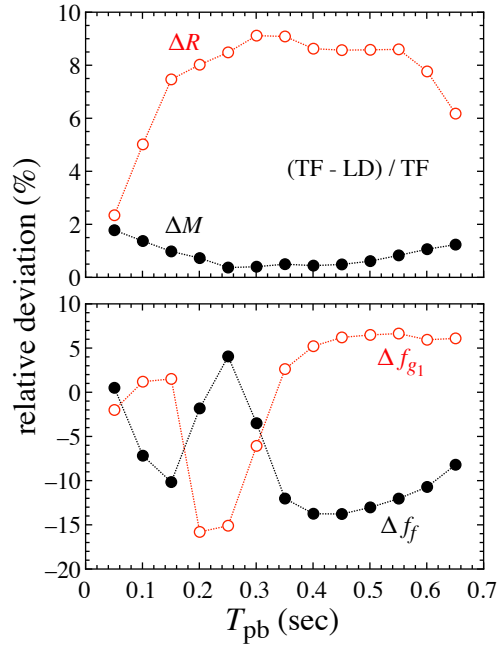} 
\end{center}
\caption{
Relative deviation of a quantity with TGLD to that with TGTF, which is calculated by $\Delta \xi = (\xi_{\rm TF}-\xi_{\rm LD})/\xi_{\rm TF}$ for some property ($\xi$) calculated with TGTF ($\xi_{\rm TF}$) and TGLD ($\xi_{\rm LD}$). The top panel corresponds to $\Delta M$ and $\Delta R$, while the bottom panel is $\Delta f_f$ and $\Delta f_{g_1}$.  
}
\label{fig:DMRft}
\end{figure}

To quantify how the evolution of observables depends on the different treatment of non-uniform matter in EOS, we calculate the relative deviation of quantity with TGLD to that with TGTF via
\begin{equation}
  \Delta \xi = (\xi_{\rm TF} - \xi_{\rm LD})/\xi_{\rm TF}, \label{eq:devi}
\end{equation}
where $\xi$ denotes the PNS mass, radius, and gravitational wave frequencies, while $\xi_{\rm TF}$ and $\xi_{\rm LD}$ denote the corresponding properties calculated with TGTF and TGLD. The obtained result is shown in Fig.~\ref{fig:DMRft}, where the relative deviation in the PNS mass and radius is shown in the top panel, while that in the $f$- and $g_1$-mode frequencies are shown in the bottom panel. From this figure, one can observe that the PNS mass weakly depends on the different treatment in EOS, but the PNS radius and gravitational wave frequencies significantly depend on it.

\section{Conclusion}
\label{sec:Conclusion}

The supernova gravitational waves must be the next frontier in the gravitational wave astronomy. To prepare for  such a crucial opportunity with watertight conditions, one has to study the supernova gravitational waves and understand the physics behind them. In this study, we primarily focus on the dependence of gravitational wave frequency on the PNS models and derive the universal relation for the gravitational wave frequency as a function of  the PNS average density, with which one may extract the evolution of the PNS average density independently of the PNS models by observing the supernova gravitational waves.
In addition, we confirm that the PNS surface gravity can be expressed as a function of the square root of the PNS average density almost independently of the PNS models. Furthermore, we also discuss how the different treatment of the non-uniform matter in EOS affects the observables. In practice, we show that the PNS mass hardly depends on the treatment of the non-uniform matter, while the PNS radius, the $f$-, and $g_1$-mode frequencies strongly depend on it.

Similar universal relations between the frequency of gravitational waves and the observables of the PNS have been proposed in several studies  \cite{TCPOF2019b,Bizouard2021,SS2019,ST2020b}.
The universal relation in Refs.~\cite{TCPOF2019b,Bizouard2021} relates the frequencies with PNS surface gravity.
This relation has been derived using more PNS models than ours.
On the other hand, as discussed in this paper, we prefer to adopt the average density and have updated the relation proposed in Ref.~\cite{ST2020b}, increasing PNS models. Our new universal relation can predict the gravitational wave frequencies not only for the later phase but also for the early phase.
A caveat is that the new universal relation would not be applicable for a black hole formation \cite{SS2019}, 
since our PNS models are calculated with phenomenological general relativity.
Numbers of the general relativistic PNS models are necessary to obtain more robust predictions.

The time evolution of mass and radius of PNS is strongly affected by entropy and electron fraction profiles \cite{ST2016,Preau21},
that are evolved with neutrino cooling processes. The evolution of them depends on the progenitor mass \cite{Warren20,Schneider20} and EOS's properties  such as incompressibility and effective mass \cite{Yasin20,Schneider20,Andersen21}.
In this paper, we have discussed the impact of non-uniform matter treatment, which changes
 neutrino's opacity via the composition of nuclei. To make our prediction of the peak frequency more accurate, we need to construct a general formula that is valid in these vast parameter spaces.

\acknowledgments


This work is supported in part by Japan Society for the Promotion of Science (JSPS) KAKENHI Grant Numbers JP17H06357, 
JP17H06364, 
JP18H01212, 
JP19KK0354, 
JP20H04753, and 
JP21H01088 
and by Pioneering Program of RIKEN
for Evolution of Matter in the Universe (r-EMU).
This research has been also supported by MEXT as ``Program for Promoting Researches on the Supercomputer Fugaku" (Toward a unified view of the universe: from large scale structures to planets, JPMXP1020200109) and JICFuS,
the National Institutes of Natural Sciences 
(NINS) program for cross-disciplinary
study (Grant Numbers 01321802 and 01311904) on Turbulence, Transport,
and Heating Dynamics in Laboratory and Solar/Astrophysical Plasmas:
``SoLaBo-X".
Numerical computations were in part carried out on Cray XC50, PC cluster and analysis server at Center for Computational Astrophysics, National Astronomical Observatory of Japan.

\appendix
\section{Comparison between the gravitational wave signals and PNS eigenfrequencies}   
\label{sec:appendix_1}

\begin{figure}[tbp]
\begin{center}
\includegraphics[scale=0.6]{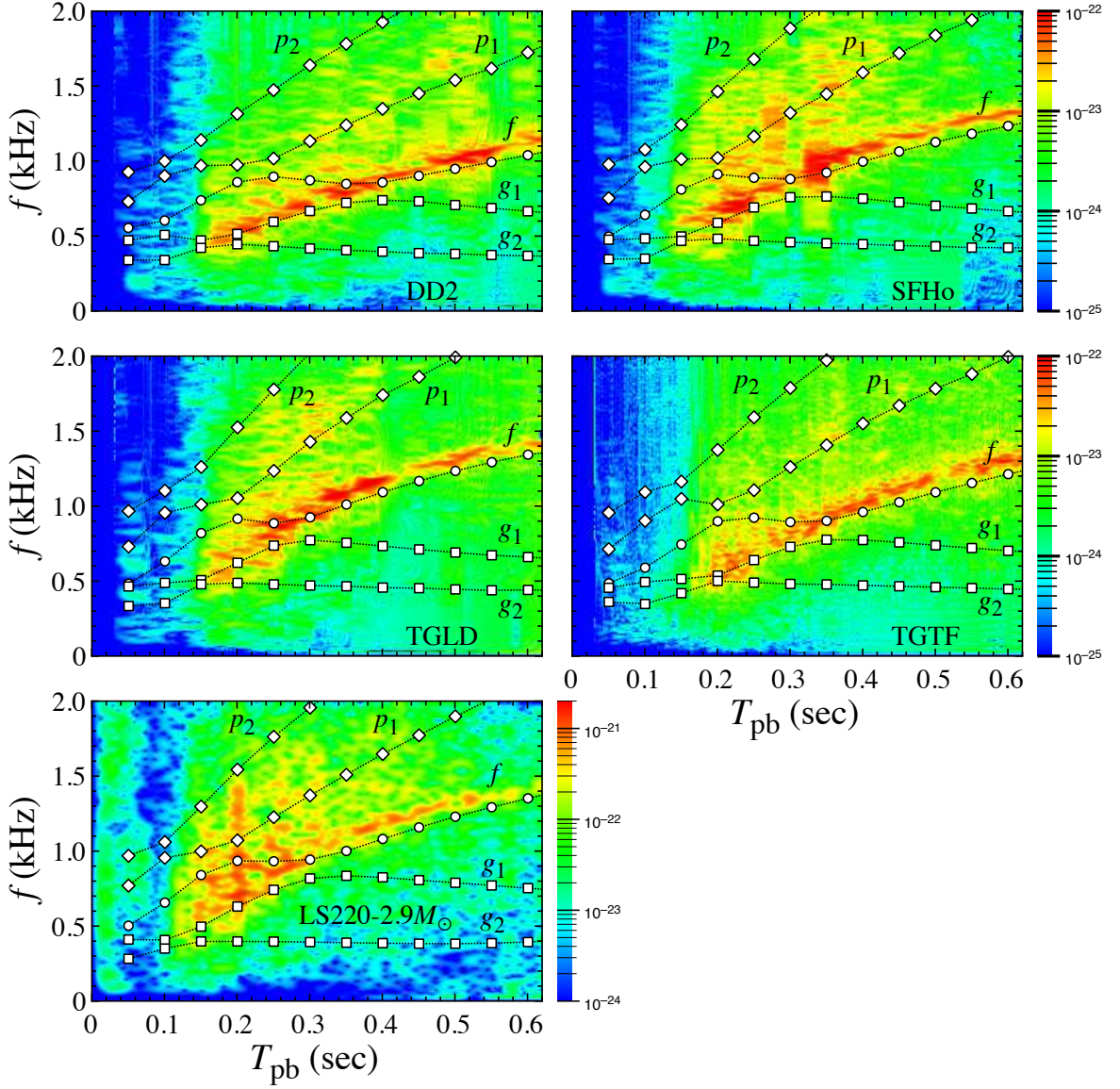} 
\end{center}
\caption{
Comparison between the gravitational wave signals appearing in the numerical simulations (contour) and the PNS specific oscillation frequencies (open marks) for various PNS models. The circles, squares, and diamonds correspond to the $f$-, $g_i$-, and $p_i$-modes with $i=1,2$, respectively. The result for the PNS model with LS220 and the $2.9M_\odot$ progenitor is taken from Ref.~\cite{ST2020b}.
}
\label{fig:comp}
\end{figure}

In the main text, we show the comparison between the gravitational wave signals appearing in the numerical simulation and the gravitational wave frequencies from the PNS only for the case with TGLD in Fig.~\ref{fig:comp-TGLD}. In Fig.~\ref{fig:comp}, we show the similar comparison for the other PNS models considering in this study. From this figure, for any PNS models, one can observe that the gravitational wave signals appearing in the numerical simulation are identified with the $g_1$-mode ($f$-mode) oscillations before (after) the avoided crossing between the $f$- and $g_1$-mode frequencies. In addition, one may be see a signature of the gravitational wave signal corresponding to the $p_1$-mode frequency in some PNS models, e.g., with SFHo, even though it is relatively weak.
In this figure, one can observe the systematical differences ($\sim 100$\,Hz) between the $f$-mode frequency and the trace in the spectrograms at late times, but unfortunately we could not identify the origin of this difference.

\end{document}